# Broadband Balanced Homodyne Detector for High-Rate (>10 Gb/s) Vacuum-Noise Quantum Random Number Generation

F. Honz, D. Milovančev, N. Vokić, C. Pacher, B. Schrenk

AIT Austrian Institute of Technology, GG4, 1210 Vienna, Austria. florian.honz@ait.ac.at

**Abstract** *We demonstrate a die-level balanced homodyne detector with a high quantum-to-classical noise clearance of 19.1dB over 3GHz bandwidth. We evaluate this receiver as a high-quality entropy source for random number generation at >10Gb/s and prove the functional integration of a classical 10Gb/s duobinary communication channel.*

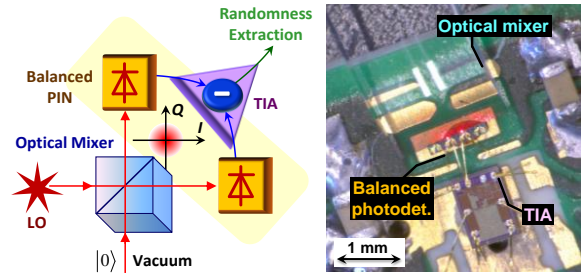

**Fig. 1:** Vacuum-noise QRNG concept and assembly.

## Introduction

Data security depends on the effectiveness of encryption methods and thus greatly relies on the availability of high-quality random numbers. The bloom of virtual services, together with the off-loading of critical tasks to the cloud infrastructure renders digital security as a key element for today's communication and information systems. Quantum random number generators (QRNG) enable such a high level of security[1] as they are based on intrinsically non-deterministic quantum physical processes.

Optical QRNGs yield truly random numbers from the measurement of optical quantum states. They are comprised of a quantum entropy source and associated optoelectronics for accessing the entropy and extracting it by means of subsequent postprocessing. Ideally, the noise due to electrical circuits is well below the quantum noise, thus quantum randomness is the dominant process. In contrast to state-of-the-art electronic-based random number generators, the entropy rate of an optical QRNG can be derived from a rather simple physical model of the device.

Optical QRNGs can be practically realized in various ways. A straightforward way is to acquire the path information of single photons; however, these QRNG engines require SPAD-based devices[2] which suffer from low photon detection probability at telecommunication wavelengths, as well as long dead-times. These limitations impose an applicability roadblock for bandwidth-critical crypto primitives such as secret sharing. Alternative approaches rely on broadband balanced homodyne detection (BHD) to measure either vacuum oscillations of light[3-5] or performing phase noise measurements[6,7]. The re-use of established componentry such as known from coherent telecommunications is also desirable in view of a seamless integration of QRNG engines into communication networks. For these reasons, QRNGs based on BHD prove as advantageous compared to their SPAD-based counterparts in terms of cost-per-bit and compactness.

In this work, we are focused on a high-speed vaccum-noise QRNG with high quantum-to-classical noise ratio (QCNR). We are leveraging on a die-level BHD receiver that features a QCNR of 19.1 dB over 3 GHz at 12.7 dBm of seed laser power. We show through experiment that random numbers can be generated at ~20 Gb/s rate and pass all NIST tests. Moreover, we experimentally prove the multi-purpose function of the QRNG engine as a classical receiver for 10 Gb/s duobinary transmission.

## Broadband BHD Receiver as QRNG Engine

The BHD in this work is comprised of die-level components arranged in a micro-optic assembly (Fig. 1), which optimises the bandwidth at the opto-electronic interface. The choice for a homodyne vacuum-noise implementation is driven by the exclusion of RF crosstalk by design. This source of EMI coupling between light source and BHD commonly applies due to laser pulsing or heterodyning in phase-noise QRNGs[8].

The die-level receiver exploits a glass-based optical mixer realized as a planar lightwave circuit

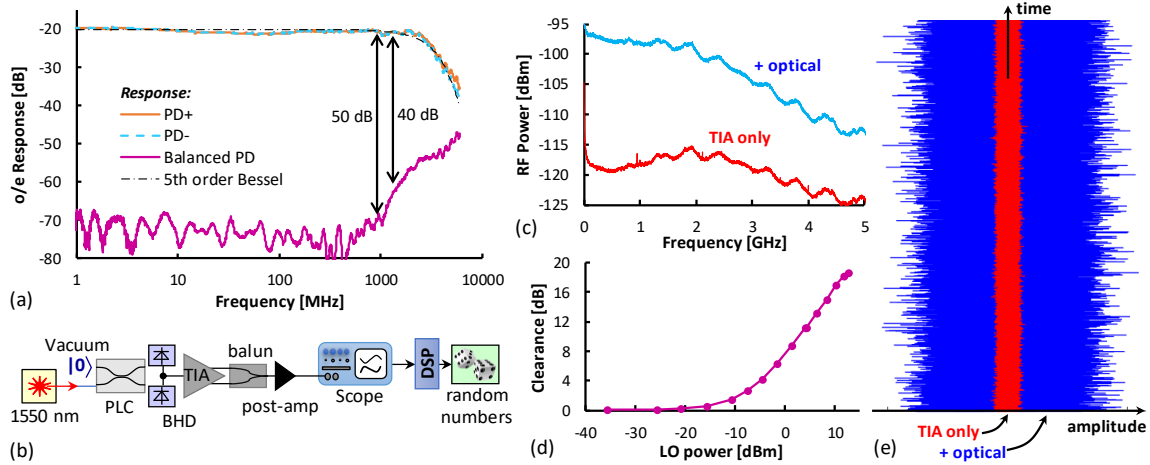

**Fig. 2:** (a) Frequency response of photodiodes and TIA, and balanced performance. (b) QRNG setup. (c) Clearance spectrum at 12.7 dBm of LO power. (d) Clearance at different LO levels. (e) Generated QRNG signal for dark and lit LO.

(PLC). After mixing local oscillator (LO) and vacuum states at an optical 180° hybrid, the light is vertically coupled to a photodiode array. Balancing of this homodyne configuration is accomplished through the assembly process, which achieves an equal power splitting without bending losses or presence of a time skew among its two optical outputs, as it generally applies to fibre-optic QRNG implementations. Moreover, the physical separation of LO and balanced photodetector ensures that the detected signal is not degraded by cladding modes of the LO, which are often a primary impediment for monolithic integrated QRNGs.

Concerning the electrical receiver front-end, a BHD receiver can be as simple as two matched photodiodes and a 50Ω termination. Although this simple construct is attractive from a bandwidth point-of-view, it leads to very high electrical noise that can jeopardize the quantum randomness. Therefore a dedicated transimpedance amplifier (TIA) is needed for converting the photocurrent into a voltage signal. In view of a high QRNG rate, a GbE TIA rated for 4.25 Gb/s is chosen. Its typical noise performance in a single-photodiode configuration is 460 nA of input-referred rms noise current and an input-reffered noise current density of 8.8 pA/√Hz. This seems relatively high at first glance, but at a wide operating bandwidth the actual distribution can still leave much room for high clearance, i.e., optical shot noise to electrical noise ratio. As long as the saturation limit is not reached for either the photodiodes or the TIA, the LO seed power can be tuned up, thus overpowering the electrical noise by a large margin.

In order to have well-matched photodiodes, a 1×4 arrayed PIN photodiode is used, where two adjacent PIN diodes can be connected with their anode and cathode as a common output of a homodyne detector to the single-ended input TIA. The use of a 1×4 array is beneficial since it allows to form a heterodyne detector configuration, which further improves the QRNG security against eavesdropper attacks arising from an unsuppressed and at the same time untrusted additional input signal[9]. Alternatively, the remaining photodiodes can be used for additional communication channels since their biasing is independent from the BHD receiver. This is further supported by the 10-GHz bandwidth property of the employed PIN photodiodes, which had a responsivity of 1.0 A/W at the target wavelength of 1550 nm.

**BHD Bandwidth and Noise Performance**

The opto-electronic bandwidth of the BHD receiver assembly is reported in Fig. 2a. The optical input has been applied to each of the two photodiodes separately, while an S-parameter measurement has been performed through LO modulation with a Mach-Zehnder modulator. The BHD receiver achieved a high bandwidth of 2.6 GHz even though the TIA is loaded with twice the individual photodiode capacitance.

The QRNG operation applies to a balanced receiver (Fig. 2b), meaning that the LO power applied to one of the mixer inputs is equally split to both photodiodes. The other input is left unconnected to measure the vacuum state.

An important aspect for balancing and a linear, non-saturated operation of the QRNG engine is an identical broadband frequency response for both photodiodes. The difference in frequency response from balanced to unbalanced case – the common mode rejection ratio (CMRR) – is 50 dB up to 1 GHz, and 40 dB at 2 GHz (Fig. 2a). Operation at an input power of -13 dBm during CMRR

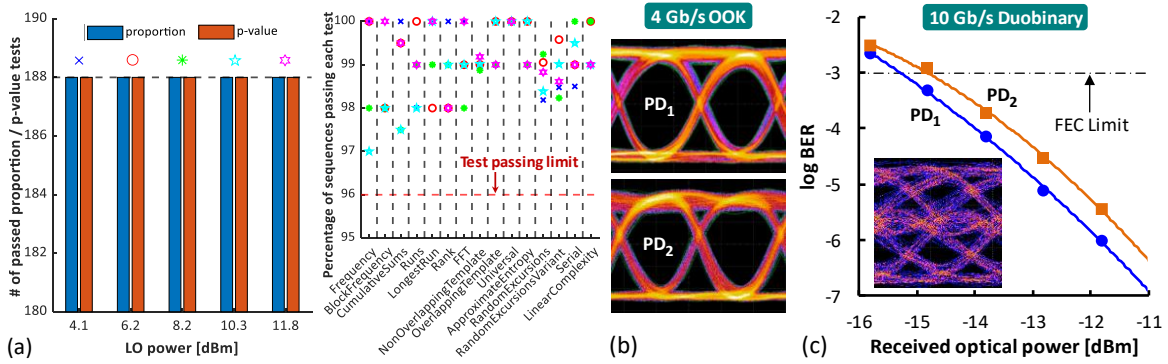

**Fig. 3:** (a) Evaluation of generated random numbers: number of tests passing the 188 NIST tests and percentage of sequences passing each proportion test. Re-use of the QRNG engine for (b) on-off keying and (c) duobinary signaling.

characterization ensured that the TIA remains linear when the BHD is unbalanced.

Regarding the noise performance, the clearance was investigated at various LO power levels. At the highest LO power of 12.7 dBm, a clearance of 20 dB is accomplished below 1 GHz (Fig 2c). It then drops, together with the TIA bandwidth. Nonetheless, a clearance of more than 10 dB is obtained up to 4 GHz; This clearance value is of interest as it resembles the minimum that is required for quantum communication applications[10].

By integrating the area under the clearance curve up to 3 GHz for a given LO power, we can get an estimation of the quantum-to-classical noise ratio over the operating bandwidth. Figure 2d shows this dependence of the clearance, integrated over a span of 3 GHz, on the LO power. The linear rise of the clearance with increasing LO power proves that the noise is quantum in nature, i.e., that we are indeed measuring the vacuum fluctuations of light. At 12.7 dBm of LO power the clearance is 19.1 dB.

Figure 2e shows the time-domain traces of TIA noise (dark QRNG) and optical seed noise (lit QRNG). The time-domain clearance can be estimated as the ratio of noise variances of the TIA outputs with (total noise) and without optical input (i.e., electrical noise). At the highest LO power of 12.7 dBm, the ratio of variances is 17.7 dB. Therefore, in frequency as well as in time domain, the balanced receiver shows a high clearance over a wide bandwidth, making it a promising candidate for fast true random number generation.

**QRNG Randomness Extraction and Testing**
For estimation of the QRNG performance we used a seeded randomness extraction algorithm[11] based on Toeplitz hashing to demonstrate the extraction of the entropy of the digitized QRNG output, which has been acquired at a resolution of 8 bits. This is necessary since the raw sampled bits are non-uniformly (Gaussian) distributed, show correlations and/or crosstalk. Finally, the complete NIST SP800-22-rev1a randomness test suite[12] has been applied and the corresponding results are reported in Fig. 3a. The generated random numbers passed all 188 tests regarding the uniformity of the p-values, as well as the proportion of sequences passing the individual tests, independent of the LO power. If we assume the lower bound of the min-entropy to be 1/4 bit per acquired bit, we estimate that we would be able to generate random numbers at a rate of ~20 Gb/s, which is equivalent to a rate of approximately $78 \times 10^6$ 256-bit AES keys per second.

**Re-Use for Classical Communications**
An important aspect for simultaneous quantum/classical communication is the use of dual-purpose physical-layer hardware. Towards this direction, we have evaluated the use of the QRNG engine for GbE signal reception, in view of the additionally available PIN diodes of the 1×4 detector array. The received signal at 4 Gb/s on-off keying shows an open eye diagram (Fig. 3b) and thus high signal integrity for both extra photodiodes. Additionally, the delivered optical power level produced a dc photocurrent of 120 µA, which was successfully both sourced and sinked by the TIA.

Since the response of the receiver (Fig. 2a) is closely following that of a 5$^{th}$-order Bessel lowpass with a -3dB cut-off of 2.5 GHz, 10 Gb/s duobinary transmission is feasible[13]. The corresponding BER performance for each of the two photodiodes is presented in Fig. 3c. The sensitivity for 10 Gb/s duobinary transmission below the FEC limit of $1 \times 10^{-3}$ is -14.8 dBm.

**Conclusion**
We have experimentally demonstrated a die-level BHD receiver with a high QCNR of 19.1 dB over a bandwidth of 3 GHz. This high noise clearance has been exploited as a high-quality (quantum) entropy source for true random number generation at >10 Gb/s rate. The generated random numbers have clearly passed all NIST tests. On top of this, we have demonstrated the multi-purpose use of the receiver assembly by integrating 10 Gb/s duobinary reception. The extension towards a quantum-heterodyne I/Q measurement is left for further investigation.

**Acknowledgements**

This work was supported in part by the ERC under the EU Horizon-2020 programme (grant n° 804769) and by the Austrian FFG Agency (grant n° 884443).